\documentclass{revtex4}

\usepackage{indentfirst}
\usepackage{amsfonts}
\usepackage{amssymb}
\usepackage{verbatim}
\usepackage{graphicx}
\usepackage{slashbox}
\usepackage[reqno,intlimits]{amsmath}
\usepackage[polish,english]{babel}

\newtheorem{theorem}{Theorem}

\begin{document}

\title{On integrable rational potentials of the Dirac equation}
\author{Tomasz Stachowiak}
\email{stachowiak@cft.edu.pl}
\affiliation{Center for Theoretical Physics PAS,\\
Al. Lotnikow 32/46, 02-668 Warszawa, Poland}

\author{Maria Przybylska}
\email{M.Przybylska@proton.if.uz.zgora.pl}
\affiliation{Institute of Physics, University of Zielona G\'ora,\\
Licealna 9, 65-417 Zielona G\'ora, Poland}

\begin{abstract}
The one-dimensional Dirac equation with a rational potential is reducible to an
ordinary differential equation with a Riccati-like coefficient. Its
integrability can be studied with the help of differential Galois theory,
although the results have to be stated with recursive relations, because in
general the equation is of Heun type. The inverse problem of finding
integrable rational potentials based on the properties of the
singular points is also presented; in particular, a general class of integrable
potentials leading to the Whittaker equation is found.

\end{abstract}

\maketitle

\section{Introduction} 


The goal of this work is to examine the solvability of the Dirac equation,
when reduced to one spatial dimension. The key point is of course that the
analysis is easier and deeper for an ordinary differential equation, and also
that the reduction itself can be achieved by many means, such as separating the
radial from the angular part, or neglecting two directions in the Cartesian
coordinates. A natural source of the potential is the magnetic field, which
also provides a direct connection with systems of quantum optics and a possible
experimental realization \cite{Martin}.
The resulting equation form is also encountered when
dealing with the supersymmetric Schr\"odinger equation \cite{Primitivo}, to which
all results on integrability can be transferred. Of course, one has to keep
track of the different definitions of potentials and physical interpretation,
but the mathematical structure remains the same, and that is the main interest
of the present Letter. 

A great many of known solutions were obtained for the scale invariant
potentials, and can be
found in \cite{Cooper}. However, such approach relies on the assumption of a
particular symmetry and is not exhaustive.
Not disregarding it, it seems natural to also consider equations like this (i.e.
second order and linear) on grounds of the differential Galois group theory.
Mostly because it provides rigorous tools for finding the exact forms of
solutions and also because it strictly rules out unsolvable cases -- exactly
like the ``ordinary'' Galois theory for polynomials.

Since linear equations often present themselves as central, like the
Schr\"odinger equation or the density perturbation in cosmology, they have
already been successfully studied with these methods, e.g. in
\cite{Primitivo,Prim2,Prim3,TS}. When it comes to the Dirac equation, the
complete analysis in the case of general polynomial potentials was carried out
in a previous work by one of the authors \cite{dirac_poly}. This class happened to be
very restrictive, and only one non-trivial case -- the so-called Dirac
oscillator -- is explicitly integrable. 

The next obvious step seems to be the wider class of rational functions which
is undertaken here. This turns out to be too wide to be
treated as completely as polynomials, the basic reason being that it is the
poles of the equation's coefficients that matter, and polynomials 
have just one -- at infinity. As will be shown, even the simplest potential
with just one pole and linear polynomial part cannot be analyzed explicitly in
that it requires recursion to formulate integrability conditions. This is a
typical feature of the Heun equation and its generalizations, and thus a
generic rational potential is beyond explicit classification which was possible
for polynomial ones. Fortunately, the recursive conditions can be studied to
some extent with regard to the real values of the parameters, and
shed some light on the number of admissible physical cases.

The analogy with polynomials also motivates the study of potentials which give
rise to similar behavior of the equation's singular points and since direct
analysis is impossible, one might also try the
inverse problem. Thus, additionally, two other scenarios are considered here:
rational potentials which produce only the irregular singularity at infinity,
and the Whittaker equation, whose solvability is known and can be used to
obtain a new, two-parameter family of integrable potentials.

\section{Sketch of the Galois theory}

To be more specific, let us consider the same setup as in \cite{dirac_poly}, so
that the time-independent Dirac equation in one dimension $z$ is
\begin{equation}
    E\psi = (-i\,\boldsymbol{\alpha}_3\,\partial_z +\boldsymbol{\beta}\, m)\psi,
\end{equation}
which gives the second order equations for each spinorial component
\begin{equation}
    \psi_{1,2}''(z) = \left(V(z)^2\pm V'(z)-E^2\right)\psi_{1,2}(z),
\end{equation}
where the Dirac matrices were taken to be
\begin{equation}
\boldsymbol{\alpha}_3 = \left(\begin{matrix} 0 & i\\ -i & 0\end{matrix}\right),\;\;
\boldsymbol{\beta} = \left(\begin{matrix} 0 & 1\\ 1 & 0\end{matrix}\right),
\end{equation}
and the potential $V$ was introduced in the so-called scalar coupling, so that
it is added to the mass term $m\rightarrow m+V$. For simplicity $m$ will
henceforth denote the effective mass thus obtained, that is the old mass plus
the constant term of $V$. The other, vector, coupling is equivalent, as shown
in \cite{dirac_poly}, so that essentially the equation in question is just
\begin{equation}
    \psi'' = (V^2+V'-E^2)\psi=:r\,\psi.
    \label{bas_eq}
\end{equation}

The question of exact solutions of such an equation with a rational coefficient
$r(z)\in \mathbb{C}(z)$ is best expressed in terms of field extensions of
$\mathbb{C}(z)$. An extension including all the elementary functions, direct
integration and some special transcendental functions is the generalized
Liouvillian extension. It consists of a finite chain of differential function
fields,
\begin{equation}
    F_0 \subset F_1 \subset \ldots \subset F_n,
\end{equation}
such that each new field $F_{k+1}$ is obtained from the previous $F_{k}$ by
adjoining an element $x_k$ which satisfies one of the three following conditions:
\begin{enumerate}
\item $x_k$ is algebraic over $F_{k}$,
\item $x_k$ is primitive, i.e. $x_k' \in F_{k}$,
\item $x_k$ is the solution of $x' = a x$, for some $a$ in $F_K$.
\end{enumerate}
These fields are differential fields precisely so that there is some notion of
the derivative $x_k'$, although here it is simply the
differentiation with regard to $z$ and the starting field $F_0$ is just
$\mathbb{C}(z)$. If the solution of a given equation belongs to $F_n$ we might
call it a Liouvillian solution and it is explicitly expressible by finite
number of elementary operations like integrals, exponentials or solving
algebraic equations. In other words, the equation in question is
solvable/integrable if it has Liouvillian solutions. 

A more detailed exposition
of the above can be found in \cite{Kaplansky}, and here it should be added
that this particular notion of exact solution is intimately connected with the
underlying group theory. Namely, one defines the differential Galois group of a
differential equation as the group of automorphisms of the field to which the
solutions belong, such that their action on the base field $F_0$ is trivial.
Intuitively it means that the group permutes the solutions, but does not change
the equation's coefficients. It
then follows that the equation is solvable when the group is solvable.
In case of the second order equations like \eqref{bas_eq} the group is an
algebraic subgroup of $\mathrm{SL}(2,\mathbb{C})$, and there exists an
algorithm contrived by Kovacic \cite{Kovacic} for deciding on its integrability
and finding the solutions (for another form of the algorithm see 
\cite{Duval}). Also, most of the common equations of mathematical
physics have been completely analyzed in this regard (see e.g. the monograph
\cite{Morales} for a collection of results), so the question is often that of
finding a connection with what is already known. This will be the main route
taken here, allowing to uncover a new, wide class of integrable potentials.

\section{Rational potentials} 

A naive generalization of the results of \cite{dirac_poly} is difficult for the
following reason.
When $V$ is polynomial, it only has one pole at infinity (and so has $r(z)$),
and it just the Laurent expansion with finitely many terms.
The polynomial's coefficients are then all we need for the analysis.
A rational $V$ can be defined by a quotient of two polynomials, and although the
denominator still has finitely many coefficients, they are not used directly.
It is the poles that matter so $r(z)$ needs to be expanded around each one. One is
therefore left with an arbitrary number of poles of arbitrary orders and coefficients
with which to construct a general rational potential that could be analyzed
effectively with Kovacic's algorithm.

Unfortunately, even the seemingly simplest case of adding just one pole
to an integrable polynomial potential leads to only implicit integrability
relations, so a full classification analogous to the polynomial case is
impossible. To see this let us consider the following
\begin{equation}
    V(z) = \frac{\alpha}{z} + m + \lambda z,
\end{equation}
with $\alpha\neq 0$. In general this yields the (biconfluent) Heun
differential equation whose solvability is determined by
algebraic relations defined only recursively -- the subject of this section.
Explicit analysis is possible whenever one of the other parameters is zero,
and the equation reduces to a special case of the confluent hypergeometric
one, which has countable families of Liouvillian solutions \cite{Morales}.

More specifically, for $\lambda=0$ Eq. \eqref{bas_eq} becomes a special
case of the Whittaker equation, analyzed in detail in the next section. When
$\lambda\neq0$ but $m=0$ it reduces to the Laguerre equation
\begin{equation}
    u g''(u) + \left(\alpha+\frac12-u\right)g'(u) = \frac{E^2}{4\lambda}g(u),
\end{equation}
where
\begin{equation}
    z^{-\alpha}\psi(z) = e^{-u/2} g(u),\;\; u = -\lambda z^2.
\end{equation}
The solutions are the Laguerre functions $L^{(\alpha-1/2)}_{n}(u)$ with
$n=\frac{-E^2}{4\lambda}$,
and these are again expressible by confluent hypergeometric functions. They are
Liouvillian if at least one of $n$ or $n+\alpha-\frac12$ is an integer
(with one solution polynomial for positive integer $n$). Additionally, $E=0$
always leads to an explicit solution of
\begin{equation}
    \psi = \exp\int V\mathrm{d}z,
\end{equation}
regardless of the form of the potential.

The generic case with all parameters non-zero can be slightly simplified with
taking $\lambda=\pm 1$ or, in other words, rescaling as follows
\begin{equation}
    z\rightarrow z/\sqrt{|\lambda|},\;\;
    m\rightarrow m\sqrt{|\lambda|},\;\;
    E\rightarrow E\sqrt{|\lambda|}.
\end{equation}
The reason for using absolute value is to keep the physical quantities real.
This will make identifying the physical integrable cases easier.

Turning now to Kovacic's algorithm, the first case will be the only one to
possible hold, and the solution must be of the form
\begin{equation}
    \psi = P(z)\exp\int\omega(z)\mathrm{d}z,
    \label{kov_sol}
\end{equation}
with rational $\omega$ and polynomial $P$. Denoting $\psi'/\psi=:\theta$ and
substituting into \eqref{bas_eq} we have
\begin{equation}
    \theta'+\theta^2 = V^2 + V' - E^2,
\end{equation}
so that the poles of $V$ and $\theta$ must agree except that there could be
additional simple poles of $\theta$ with residue 1 not appearing in $V$.
Since $\theta = \frac{P'}{P} + \omega$,
the first term is needed precisely to account for those poles, and $\omega$ has
to have the same polynomial part as $V$ (up to a sign) so that the expansions at
infinity agree. Finally, the pole $\alpha/z$ requires that $\omega$ includes
either $\alpha/z$ or $(1-\alpha)/z$, so that there are four possibilities all
in all:
\begin{equation}
    \omega_{1,2} = \frac{\alpha}{z}\pm(m+\lambda z),\;\;
    \omega_{3,4} = \frac{1-\alpha}{z}\mp(m+\lambda z).
\end{equation}

The next step is to substitute the above into the equation in question and look
for polynomial solutions for $P$ with its degree $d$ determined by the highest
order term in each respective equation. The four subcases are thus
\begin{equation}
\begin{aligned}
    1.&& P'' + 2\left( \frac{\alpha}{z}+m+\lambda z\right)P' &= 2\lambda d P,&
    d &= -\frac{E^2}{2\lambda}\\
    2.&& P'' + 2\left( \frac{\alpha}{z}-m-\lambda z\right)P' &= 
           \left(\frac{4\alpha m}{z}-2\lambda d\right)P,&
    d &= \frac{E^2}{2\lambda} -2\alpha -1,\\
    3.&& P'' + 2\left( \frac{1-\alpha}{z}-m-\lambda z\right)P' &= 
           \left(\frac{2 m}{z}-2\lambda d\right)P,&
    d &= \frac{E^2}{2\lambda} -2,\\
    4.&& P'' + 2\left( \frac{1-\alpha}{z}+m+\lambda z\right)P' &= 
           \left(\frac{2 m(2\alpha-1)}{z}+2\lambda d\right)P,&
    d &= -\frac{E^2}{2\lambda} + 2\alpha-1.
    \label{energ}
\end{aligned}
\end{equation}
Taking
\begin{equation}
    P = \sum_{l=0}^{d}p_l z^l,
\end{equation}
the above can be stated as recurrence relations
\begin{equation}
\begin{aligned}
    1.&& l(l+2\alpha-1)p_l + 2m(l-1)p_{l-1} &= 2\lambda(d-l+2)p_{l-2},\\    
    2.&& l(l+2\alpha-1)p_l - 2m(l+2\alpha-1)p_{l-1} &=2\lambda(l-d-2)p_{l-2},\\
    3.&& l(l+1-2\alpha)p_l - 2mlp_{l-1} &= 2\lambda(l-d-2)p_{l-2},\\
    4.&& l(l+1-2\alpha)p_l + 2m(l-2\alpha)p_{l-1} &= 2\lambda(d-l+2)p_{l-2},
    \label{rec_rel}
\end{aligned}
\end{equation}
with common initial conditions of
\begin{equation}
    p_{d+1} = 0,\;\; p_{d}=1.
\end{equation}
They reflect the fact that $P$ can always be taken to be a monic polynomial,
and its degree is exactly $d$. Consequently, the relations hold for
$l\leq d+1$ and one can successively generate $p_l$, as functions of the
parameters, down to $p_0$. The last but one relation, i.e. for $l=2$ gives us
$p_0$ but there is one more with $l=1$ and only two terms $p_1$ and $p_0$ since
for polynomials $p_{-1}\equiv 0$. This is the required condition for the
existence of polynomial $P$. It is convenient, however, to consider the last
relation as the definition of $p_{-1}$, which will be a function of $m$,
$\lambda$, $\alpha$ and $d$, and then the condition $p_{-1}=0$ (equivalent to
the two-term relation between $p_1$ and $p_0$) will yield the
admissible values of parameters for which a polynomial $P$ exists, and hence
the main equation is solvable. Since the whole problem is considered over the
complex domain, all polynomials have the maximal number of roots, counting with
multiplicities. When considered as polynomials in $m$, $p_{-1}$ is of degree
$d+1$ provided that the middle term of the recurrence does not vanish. This
particular way that $m$ enters also means that all the polynomials are either
even or odd functions of $m$.

For each degree $d$ the relations \eqref{rec_rel} implicitly define $p_{-1}$
whose roots then provide the value of parameters leading to an integrable case,
while the energy is determined by its relation with $d$ in each case.
Of course, one is mostly interested in the real values of the parameters as
they correspond to the physical case. This is the reason for approaching $p_l$
as polynomials in $m$ and $\alpha$ ($\lambda$ can be scaled away, as mentioned
before). The particular form of the recurrence makes those sequences define
orthogonal polynomials in most cases (although this property is of no use
here), and their zeros are then all real, simple and with the separation
property
\begin{equation}
    a_{i}^{(k+1)} < a_{i+1}^{(k)} < a_{i+1}^{(k+1)},
\end{equation}
where $a_i^{(k)}$ is the $i$-th zero of the $k$-th polynomial.
To see this in the first subcase note that 
\begin{equation}
    p_{d-1} = -dm,\;\; a_1^{(d-1)} = 0, 
\end{equation}
and
\begin{equation}
    p_{d-2} = d(2(d-1)m^2+1-2\alpha-d)/4,\;\;\;\;
    a_1^{(d-2)} = -\sqrt{\frac{d+2\alpha-1}{2(d-1)}},\;
    a_2^{(d-2)} = \sqrt{\frac{d+2\alpha-1}{2(d-1)}},
\end{equation}
so that distinct real roots exist for $2\alpha > 1-d$. $d=0$ is ruled out by
$E\neq 0$, $d=1$ ruled out by $m\neq 0$, so the smallest $d$ is 2, which
translates to $\alpha > -1/2$. Assuming that the
separation property holds for polynomials $p_l$ and $p_{l-1}$ the proof is by
induction on $l$. Evaluating $p_{l-2}$ at a zero of $p_{l-1}$ gives
\begin{equation}
    l(l+2\alpha-1)p_l(a_i^{(l-1)}) = -2(d-l+2)p_{l-2}(a_i^{(l-1)}),
\end{equation}
where $\lambda$ was taken negative, as is necessary by the energy condition in
\eqref{energ}. Since the zeros of $p_l$ lie only between those of $p_{l-1}$ it
follows that the left-hand side of the above is non-zero and switches sign as
$i$ increases, provided that $l+2\alpha-1$ is non-zero. It follows that
$p_{l-2}$ must change sign in each interval $(a_i^{(l-1)},a_{i+1}^{(l-1)})$
which can happen only at a root. The only thing left to show that $p_{l-2}$ has
two roots outside of $(a_1^{(l-1)},a_{d-l+1}^{(l-1)})$. With the further
assumption that $l+2\alpha-1>0$ the sign of $p_{l-2}$ at the last root of
$p_{l-1}$ is opposite to the sign of $p_{l}$ but they both have the same sign
in their leading powers of $m$ which are $m^{d+1}$ and $m^{d-1}$ respectively,
so that $p_{l-2}$ must have another root for the
signs to agree as $m\rightarrow\infty$. The same happens at the lowest root.
Finally, care has to be taken as to the middle term of the recurrence, as it
cannot vanish (the number of roots must be maximal). Fortunately for $l=1$ the
relation just states that 
$\alpha p_1 = 2\lambda(d+1)p_{-1}$ so effectively we can take $l\geq 3$ and
$p_1=0$ as the integrability condition, and automatically the smallest $d$ to
consider is 3 as $d=2$ implies $p_{-1}=-2m$. Altogether this means that the
induction is valid for $\alpha >-1$, and the separation property holds up to
$p_1$.

The third subcase is analogous and even simpler, with the only
difference being
that the energy condition requires $\lambda>0$, and the positivity of 
$l+2\alpha-1$ is enough to guarantee the appropriate changes of signs, and
maximal degree of each $p_l$. For the first induction step, the condition for
$p_{d-2}$ to have simple real roots is that $d-2\alpha+1$ be positive. Since
$d=0$ gives $p_{-1}=m$, it follows that $\alpha > 1$ is enough for the
separation property.

Subcases two and four differ in that the sign of $\lambda$ is now arbitrary,
but the general proofs are the same and we get $\lambda >0$, $\alpha >0$ or
$\lambda <0$, $\alpha<-d/2$ in subcase two, and $\lambda >0$, $\alpha >(d+1)/2$
or $\lambda<0$, $\alpha < 1/2$. Where again the trivial cases $m=0$ were
discarded. However, the results can be slightly strengthened here. In subcase
two, for $ \alpha =0$ the relation with $l=1$ just gives $p_{-1}\equiv 0$, and
one can see that for $-1/2 < \alpha < 0$, the induction can still be carried
as before, with the modification that signs at the last step are reverted.
Namely 
\begin{equation}
    2\alpha p_1 -4m\alpha p_0 = -2\lambda(d+1)p_{-1},
\end{equation}
so that for $\lambda>0$ at the maximal root of $p_0$ the signs of $p_{-1}$ and
$p_1$ are the
same, but additionally the leading coefficients differ in sign, as the
coefficient multiplying $m p_0$ in the relation is positive, while it was
negative for all higher $l$. Thus the existence of the highest root follows.
For $\lambda<0$ this cannot be done, as the main restriction comes from the
roots of $p_{d-2}$ not from the third induction step.

The same is true for subcase four with $(d+1)/2 > \alpha > d/2$ and $\lambda>0$
so that the region where the separation property holds can be extended to
$\alpha>d/2$. The case of $\alpha=(d+1)/2$ is, by the energy condition,
equivalent to $E=0$ so it brings no new solutions.

Putting the above together, and discarding the $m=0$ case we have
\begin{theorem}
The integrability condition $p_{-1}=0$, considered as an equation in $m$ has
the full number of single real roots in the subcases \eqref{rec_rel} for
\begin{enumerate}
\item $\lambda<0$ and $\alpha>-1$. There are $\lfloor (d-1)/2\rfloor$ positive
    roots with $d = -\frac{E^2}{2\lambda}$.
\item $\lambda>0$ and $\alpha>-\frac12$, or $\lambda<0$ and $\alpha<-d/2$. There
    are $\lceil d/2\rceil$ positive roots with
    $d=\frac{E^2}{2\lambda}-2\alpha-1$.
\item $\lambda>0$ and $\alpha < 1$. There are $\lceil d/2\rceil$ positive roots
    with $d=\frac{E^2}{2\lambda}$.
\item $\lambda>0$ and $\alpha>d/2$, or $\lambda<0$ and $\alpha<\frac12$. There
    are $\lceil d/2\rceil$ positive roots with
    $d=-\frac{E^2}{2\lambda}+2\alpha-1$.
\end{enumerate}
\label{th1}
\end{theorem}
Recall that $\lfloor\;\rfloor$ and $\lceil\;\rceil$ are the floor and ceiling
functions, respectively.

Next, the inspection of the $m=0$ case tells us that for odd $d=2n+1$ the
conditions read
\begin{equation}
\begin{aligned}
    1.&& p_{-1} &= 
    \prod_{n=0}^{\frac12 (d-1)}\frac{(2n+1)(\alpha+n)}{\lambda(d-2n+1)}&=0,\\
    2.&& p_{-1} &= 
    \prod_{n=0}^{\frac12 (d-1)}\frac{(2n+1)(\alpha+n)}{\lambda(2n-d+3)}&=0,\\
    3.&& p_{-1} &= 
    \prod_{n=0}^{\frac12 (d-1)}\frac{(2n+1)(n+1-\alpha)}{\lambda(2n-d+3)}&=0,\\
    4.&& p_{-1} &= 
    \prod_{n=0}^{\frac12 (d-1)}\frac{(2n+1)(n+1-\alpha)}{\lambda(d-2n+1)}&=0,
    \label{odd_pprod}
\end{aligned}
\end{equation}
so that there are always $n=(d-1)/2$ real simple roots. This happens because
the recurrence involves only two terms then, and we start with $p_d$ reducing
the index by 2 at each iteration to get to $p_{-1}$. It then follows that in
some neighborhood of $m=0$ there must still be $n$ real simple roots, because
$p_{-1}$ is a polynomial in $\alpha$ whose coefficients are again polynomials
of $m$, so that the roots must be continuous functions of $m$. Note however,
that in subcases 2 and 4, the full polynomial could be of higher degree in
$\alpha$, thanks to the middle term, and there will be additional roots, that
go to infinity when $m\rightarrow 0$.

The analysis of even degrees is much more involved, so we include the proof in
the appendix, and just state the general result here

\begin{theorem}
    In the vicinity of $m=0$ there is always the maximal number of real simple
    roots in $\alpha$ of the integrability condition $p_{-1}=0$. Depending on
    the parity of $d$ and the subcases of \eqref{rec_rel} one has
\begin{enumerate}
\item  For odd $d$ there are $(d+1)/2$ roots (including $\alpha=0$)
    given by \eqref{odd_pprod}. For even $d$ there are $d/2$ roots interleaved
    with the roots of \eqref{p_even1}.
\item  For odd $d$ there are $(d+1)/2$ roots (including $\alpha=0$)
    given by \eqref{odd_pprod}. For even $d$ there are $d/2$ roots interleaved
    with the roots of \eqref{p_even2} (also including $\alpha=0$).
\item  For odd $d$ there are $(d+1)/2$ roots 
    given by \eqref{odd_pprod}. For even $d$ there are $d/2$ roots interleaved
    with the roots of \eqref{p_even3}.
\item  For odd $d$ there are $(d+1)/2$ roots, one of them being 
    $\alpha=(d+1)/2$, given by \eqref{odd_pprod}. For even $d$ there are $d/2$
    roots interleaved with the roots of \eqref{p_even4}, except for the the
    shared root $\alpha=(d+1)/2$.
\end{enumerate}
\label{th2}
\end{theorem}

\begin{figure}[h]
    \includegraphics[width=.4\textwidth]{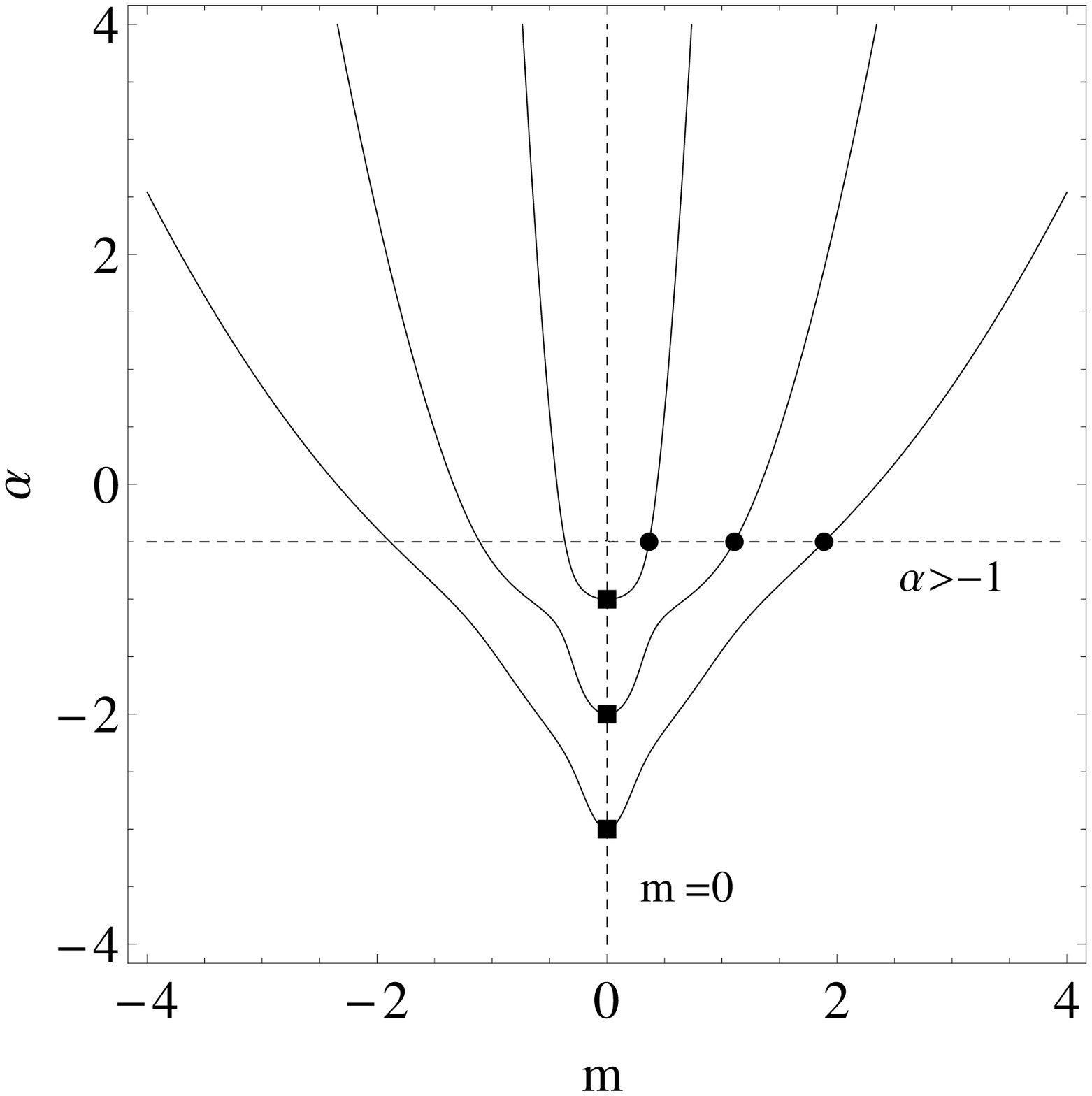}
    \includegraphics[width=.4\textwidth]{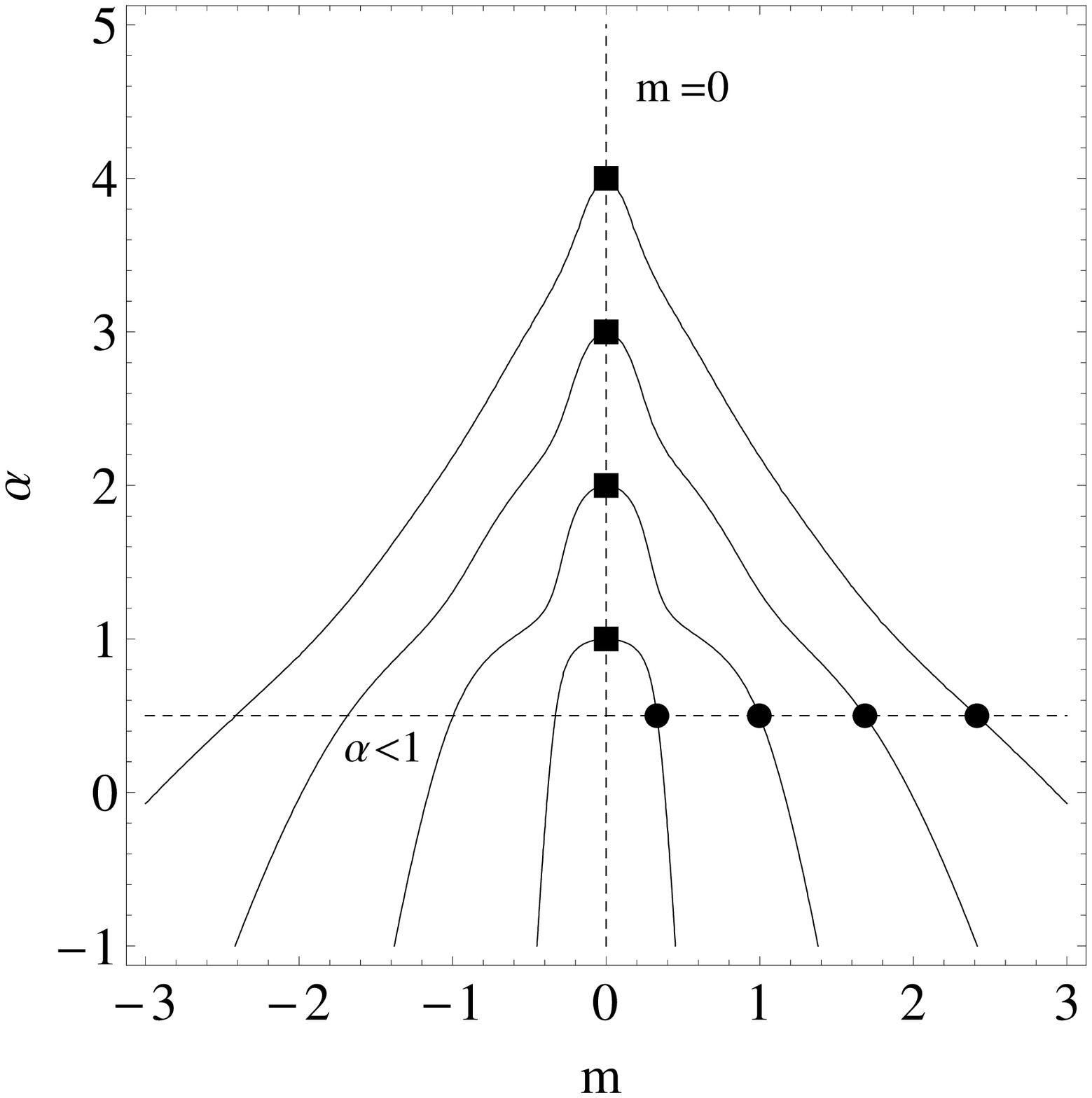}
    \caption{ The curves (solid) defined by $p_{1}=0$ in subcases 1 and $p_{-1}=0$ in
    subcase 3, both with $d=7$. Circles represent zeroes obtained with the separation
property, while squares are zeroes for $m=0$ that change into simple
roots for small positive (and negative) $m$.}
    \label{gr13}
\end{figure}

\begin{figure}[h]
    \includegraphics[width=.4\textwidth]{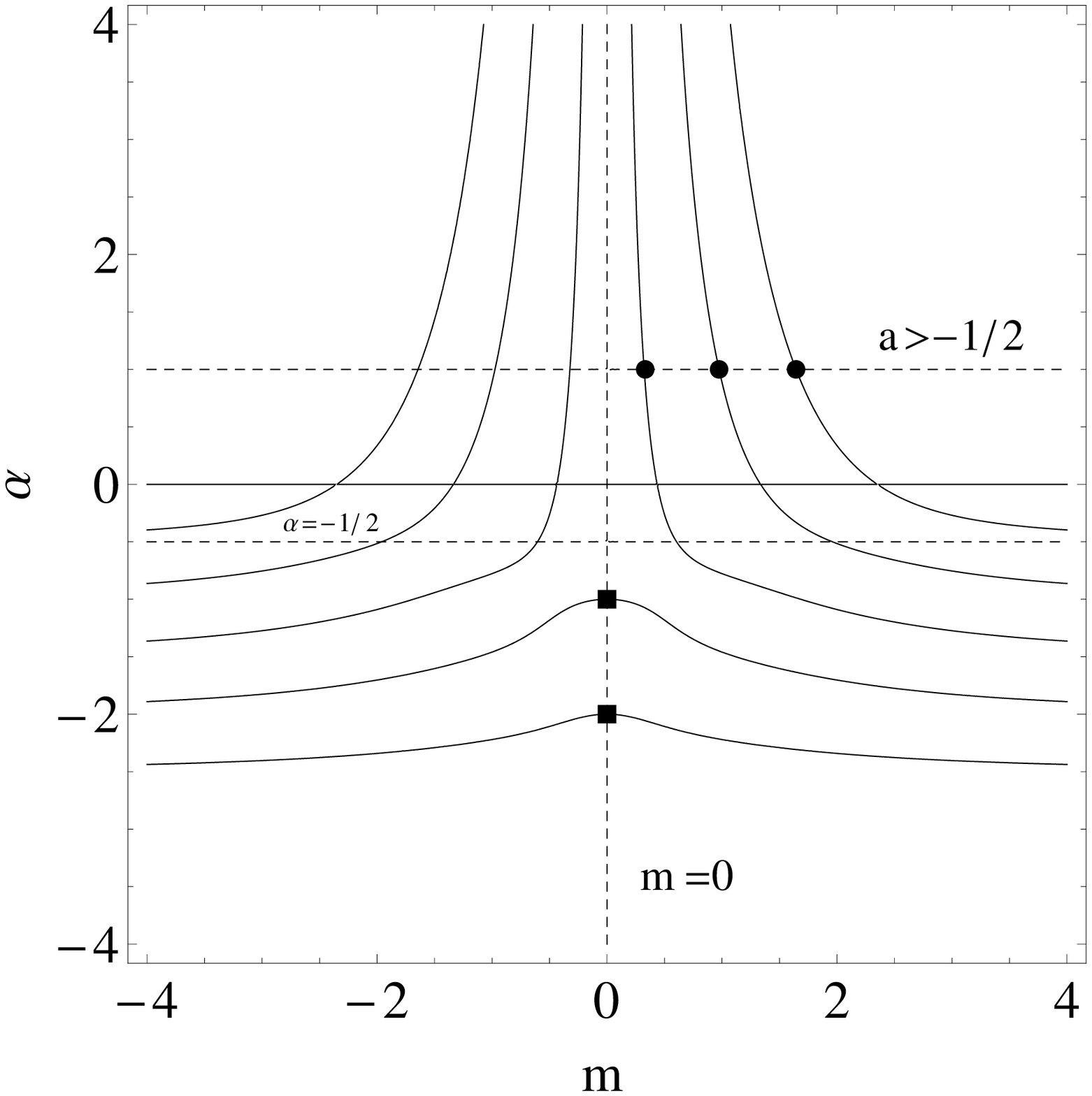}
    \includegraphics[width=.4\textwidth]{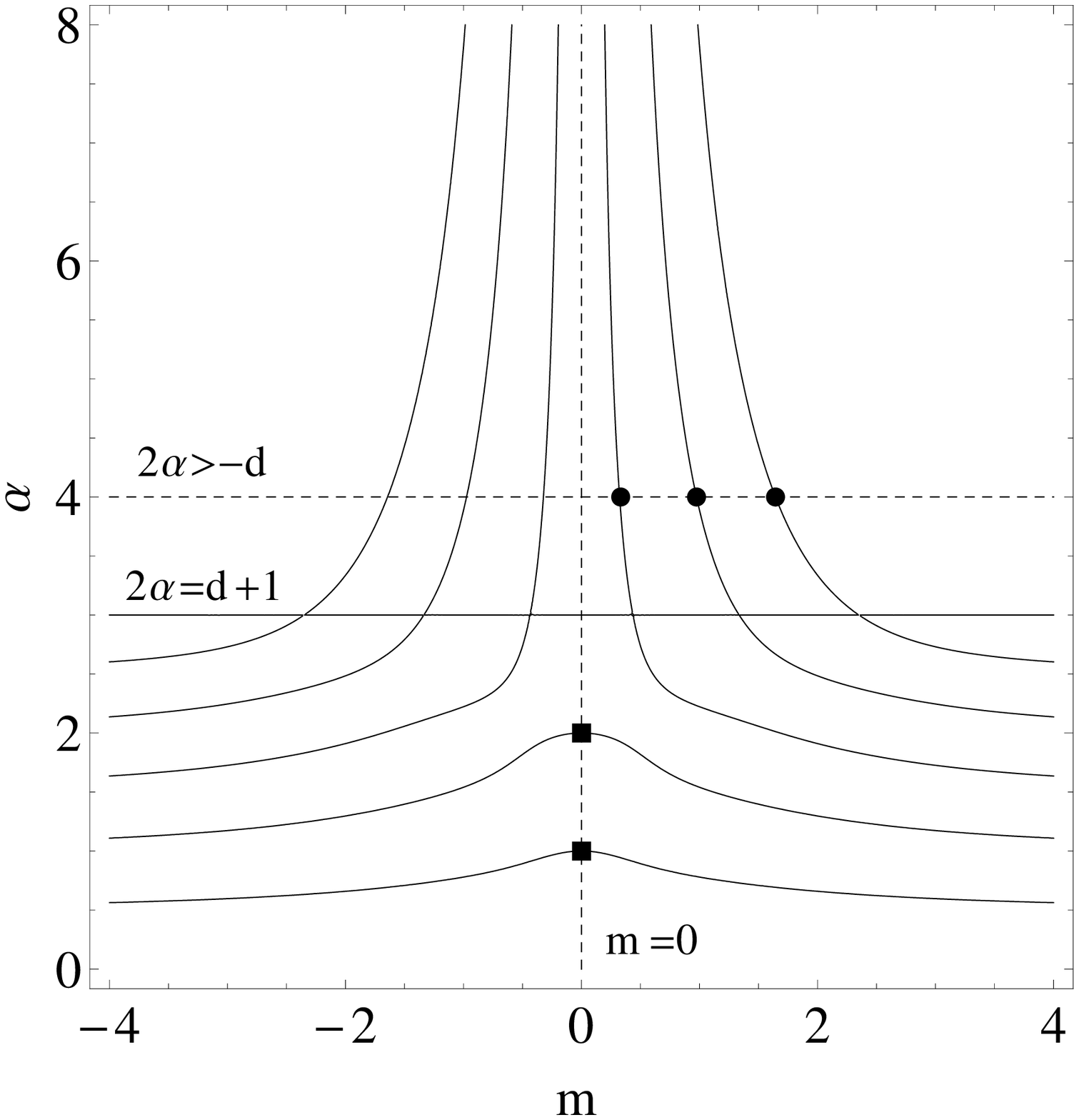}
    \caption{ The curves (solid) defined by $p_{-1}=0$ in
    subcases 2 and 4, both with $d=5$. Circles represent zeroes obtained with
    the separation property, while squares are zeroes for $m=0$ that change
    into simple roots for small positive (and negative) $m$.}
    \label{gr24}
\end{figure}

The the best way to view all the branches of the algebraic sets defined by the
integrability conditions, and to get the intuition about the asymptotic
behavior of roots, is a plot of $p_{-1}=0$ in the $(m,\alpha)$ plane shown in
Figs. \ref{gr13} and \ref{gr24}.

The graphs suggest that there are horizontal asymptotes in
subcases 2 and 4, so
that for large $m$ there would also need to exist a fixed number of solutions.
To check this, let us take rescaled polynomials defined by $p_l = m^{d-l} r_l$
which they satisfy
\begin{equation}
\begin{aligned}
    2.&& l(l+2\alpha-1)\frac{r_l}{m^2} - 2(l+2\alpha-1)r_{l-1} &=2\lambda(l-d-2)r_{l-2},\\
    4.&& l(l+1-2\alpha)\frac{r_l}{m^2} + 2(l-2\alpha)r_{l-1} &=
    2\lambda(d-l+2)r_{l-2},
\end{aligned}
\end{equation}
where the rescaling is based on the maximal power of $m$ appearing in $p_l$.
This means that each $r_l$ is finite as $m\rightarrow\infty$ and the
limit case of the relations reads
\begin{equation}
\begin{aligned}
    2.&& -2(l+2\alpha-1)r_{l-1} &=2\lambda(l-d-2)r_{l-2},\\
    4.&& 2(l-2\alpha)r_{l-1} &= 2\lambda(d-l+2)r_{l-2},
\end{aligned}
\end{equation}
so that at each recurrence step a real zero in $\alpha$ is introduced --
exactly as in the $m=0$ case above. In
subcase 2 $l=1$ adds the precluded $\alpha=0$, while in subcase 4, $l=d+1$ is
the previously mentioned straight line of solutions (independent of $m$). These
real roots at infinity cannot turn into complex one for large finite $m$ since
any complex roots would have to come in conjugate pairs. This would require the
polynomials of twice the degree, but also the roots at infinity to be double.
By continuity we thus obtain
\begin{theorem}
    For sufficiently large $m^2$ there are always solutions of $p_{-1}=0$, when
    considered as an equation in $\alpha$. In subcase 2, there are $d$ solutions
    which asymptotically approach $\alpha=(1-l)/2$ for $l=2,\ldots,d+1$. In
    subcase 4, there are $d$ solutions that approach $\alpha = l/2$ for
    $l=1,\ldots,d$ and the line $\alpha=(d+1)/2$ is also always a
    solution.
    \label{th3}
\end{theorem}

The question remains as to the vertical asymptotics of the branches, that is
what happens for infinite $\alpha$. Rescaling this time by
$p_l=\alpha^{d-l}r_l$ gives
\begin{equation}
\begin{aligned}
    2.&& -4mr_{l-1} &=2\lambda(l-d-2)r_{l-2},\\
    4.&& -4mr_{l-1} &= 2\lambda(d-l+2)r_{l-2},
\end{aligned}
\end{equation}
so that only the factor of $m$ is introduced at each step. This means that all the
branches tend to $m=0$. For any straight line of constant $m$, each branch must
intersect it then, as the previous theorem guarantees that they extend towards
infinite $m$.
If in the vicinity of $m=0$ it were possible to prove the existence of
solutions also for odd degrees, that would translate into global existence of
exactly $d+1$ roots in $\alpha$ for any value of $m$ (counting also $\alpha=0$
and $\alpha=(d+1)/2$ where necessary).

Finally, the last two cases of Kovacic's algorithm need to be dealt with. The
second case cannot hold for the required quantities $E_c$ as defined in
\cite{Kovacic} are: $E_0 = \{4(1-\alpha),2,4\alpha\}$ and $E_{\infty}=\{-2\}$.
Then the degree of the polynomial $P$ defined as above should be 
$d=(e_{\infty} - e_0)/2$ for $e_x\in E_x$, and this implies that $\alpha$ has
to be an integer or half integer. However this makes all $e_x$ even, so that
case two cannot hold. Case three is ruled out because the order of $r(z)$ at
infinity is less than 2. In other words infinity is an irregular singular
point.

\section{Reconstruction of the potential}
The direct approach above gave us only implicit conditions and with so many
general classes of integrable equations it is natural to try and work
ones way in the other direction, i.e. reconstructing the integrable potentials.
To demonstrate the idea let us first consider how to generalize the basic form
of the equation in question.

We want our main equation \eqref{bas_eq}
to assume some particular, prescribed form with rational $r(z)$, and notice
immediately that energy enters in such a way that the infinity must be an
irregular singular point. This could in principle be remedied by other
parameters of $V$, but only for one fixed value of $E^2$, while the physical
nature of the problem allows any $E$ to begin with (and usually and with a
restricted, but still infinite spectrum). It is thus better to assume
that infinity is irregular, and the special values of parameters for which it
is not will up in the analysis anyway.

At first one might ask whether there are rational potentials with the property
that only the infinity is a singular point -- like in the polynomial case.
$V(z)$ would then need at least one pole, not to be polynomial, and it could
only be of first order and residue 1 if $V^2+V'$ is to have no poles. The
general potential would then be
\begin{equation}
    V(z) = \sum_{k=1}^{N} \frac{1}{z-c_k} + R(z),
\end{equation}
with a polynomial $R(z)$, and $r(z)$ reads
\begin{equation}
    r(z) = R^2+R'+ 2R\sum_{k=1}^N \frac{1}{z-c_k} + 
    2\sum_{\substack{k,l=1\\k\neq l}}^N \frac{1}{(x-c_k)(x-c_l)}-E^2.
\end{equation}
The only possibility of removing the poles is cancellation between the two sums,
and expanding them at each $c_k$ the condition is
\begin{equation}
    R(c_k) + \sum_{\substack{l=1\\l\neq k}}^N \frac{1}{c_k-c_l} = 0.
    \label{R_cond}
\end{equation}
The poles are all different (otherwise the residue would not be 1), so that
this system of $N$ equations determines a polynomial of degree $N-1$ uniquely,
but any other polynomial which vanishes at the poles can be added, so the most
general form of $R$ is
\begin{equation}
    R(z) = R_1(z) + R_2(z)\prod_{k=1}^N(z-c_k),
\end{equation}
where $R_1$ satisfies \eqref{R_cond} and $R_2$ is arbitrary. The system
determining coefficients of $R_1$ has no explicit general solution (inverse of
the Vandermond matrix) for arbitrary $N$, so let us again look at the potential
with just one pole at zero so that
\begin{equation}
    R(z) = z R_2(z) = z\sum_{k=0}^n r_k z^k,
\end{equation}
and
\begin{equation}
    r(z) = 3R_2 + zR_2'+z^2R_2^2-E^2.
\end{equation}
With no poles in $r(z)$ and irregular infinity, only the first case of Kovacic
algorithm can hold, so that the solution must be of the form \eqref{kov_sol}.
Substituting this into the differential equation \eqref{bas_eq} and comparing
with $r(z)$ at
infinity tells us that $\omega = zR_2$ and also that the degree of $P$ must
be 1. Taking the polynomial to be monic $P(z)=z+p_0$ results in the final
equation
\begin{equation}
    (E^2-2r_2)p_0 +E^2 z = 0,
\end{equation}
which can hold only when $R_2$ is of degree 1 with $p_0 = r_0/r_1$ and $E^2 =
2r_0$ ($r_1=0$ leads to $E=0$ again). The integrable potential is then
\begin{equation}
    V(z) = \frac{1}{z} + z(r_0+r_1 z),
\end{equation}
and the solutions
\begin{equation}
\begin{aligned}
    \psi_1(z) &= (r_1 z+ r_0) \exp\left(z^2(3r_0+2r_1z)/6\right),\\
    \psi_2(z) &= \psi_1 \int\frac{1}{\psi_1(z)^2}\mathrm{d}z.
\end{aligned}
\end{equation}

For higher $N$ it is possible to proceed halfway in that the expansion of
$r(z)$ at infinity implies that $\omega = R(z)$, and the solution of the
prescribed form $P\exp\int\omega$ leads to
\begin{equation}
    P'' + 2R P'+(R^2+R'-r)P = 0.
\end{equation}
The highest order term vanishes when deg$P=N$, so this is just a linear system
for the coefficients of $P(z)$, but to write it out explicitly one would have
to know $R_1$, which amounts to solving \eqref{R_cond} separately for each
value of $N$.

\section{Whittaker equation}

Finally, let us look at a known type of equation for which non-trivial
integrable cases exist. In contrast to the previous example, we expect $r(z)$ to
have a pole but want to restrict the degree at infinity (the polynomial part).
As mentioned before, there is a special case of Heun
equation which reduces to the confluent hypergeometric one and has a finite
pole of order two in $r(z)$, while only the energy accounts for infinity being
irregular. Its general form is the Whittaker equation:
\begin{equation}
    y'' = \left(\frac14 -\frac{\kappa}{z} +\frac{4\mu^2 -1}{4z^2}\right)y,
\end{equation}
and since its solvability conditions are known, we can find out which potentials
allow Liouvillian solutions.

In order for the Dirac equation to assume such form, the potential has to
satisfy
\begin{equation}
    V^2+V'-E^2 = 
    \left(\frac14 -\frac{\kappa}{z} +\frac{4\mu^2 -1}{4z^2}\right),
\end{equation}
which is a Riccati equation. Its rational solution can be obtained by the
analysis of the poles of the right-hand side. The general solution to the above
is
\begin{equation}
    V = m + \frac{\alpha}{z} +\sum_{k=1}^{d}\frac{1}{z-c_k}.
\end{equation}
The poles $c_k$ have to be included even though they do not correspond to the
poles of $V^2+V'$ because each term of the sum is a solution to $V^2+V'=0$.
Since the pole at zero is essential in the
equation, its residue is $\alpha\neq1$. Alternatively, the above could be
rewritten as
\begin{equation}
    V = m + \frac{\alpha}{z}+\frac{P'(z)}{P(z)},
\end{equation}
where $P$ is a polynomial with $d$ distinct roots $c_k$, such that $P(0)\neq0$,
and can be taken to be monic.

Substituting this form of $V$ to the coefficient $r$, one gets
\begin{equation}
    r = m^2 -E^2 +\frac{\alpha(\alpha-1)}{z^2} + \frac{2m\alpha}{z} +
    \frac{2(m z+\alpha)P'+z P''}{z P}.
\end{equation}
The restrictions on the singular points require that the last term is of the
form $a+b/z+c/z^2$, so that we have an equation for $P$
\begin{equation}
    zP'' + 2(\alpha+m z)P' = \left(a z + b +\frac{c}{z}\right)P.
\end{equation}
As the solution is supposed to be a polynomial, the coefficient $a$ must be
zero, or the terms of the highest order would not cancel. Similarly, the lowest
order terms require that $P(0)c=0$ and since we excluded zero from the list of
the roots $c$ also must vanish. We now have
\begin{equation}
    zP'' + 2(\alpha+m z)P' = b P,
\end{equation}
which is the associated Laguerre equation, with polynomial solutions only for 
\begin{equation}
b = 2m d,\;\; d\in \mathbb{N},\; m\neq 0,
\end{equation}
so that
\begin{equation}
    P = L_d^{2\alpha-1}(-2mz).
\end{equation}

With such condition, the Dirac equation becomes
\begin{equation}
    \psi'' = \left(m^2-E^2 + \frac{\alpha(\alpha-1)}{z^2} +
    \frac{2m(\alpha+d)}{z}\right)\psi,
    \label{pre_whitt}
\end{equation}
and assuming $m^2\neq E^2$ (the special case will be dealt with below)
a change of the dependent variable to
\begin{equation}
u = 2\sqrt{m^2-E^2}\, z
\end{equation}
gives
\begin{equation}
    \psi''(u) = \left(\frac14 + \frac{\alpha(\alpha-1)}{u^2} +
    \frac{m}{\sqrt{m^2-E^2}}\frac{\alpha+d}{u}\right)\psi,
\end{equation}
so that the appropriate Whittaker's equation's coefficients are
\begin{equation}
\begin{aligned}
    \mu &= -(\alpha-\frac12),\\
    \kappa &= -\frac{m}{\sqrt{m^2-E^2}}(\alpha+d),
\end{aligned}
\end{equation}
and we simply have
\begin{equation}
    \psi(z) = W_{\kappa,\mu}\left(2\sqrt{m^2-E^2} z\right).
\end{equation}

Since the equation is of the second order, there will be also another, linearly
independent solution given by $W_{-\kappa,\mu}(-z)$.
However, in application one will choose only one taking into account the
physical context such as square integrability, and the domain of $z$. For
example, in the  $z\gg0$ region, one would not take the function
$W_{-\kappa,\mu}(-z)$ since it behaves as $\exp(z)$ asymptotically, while
$W_{\kappa,\mu}(z)$ behaves as $\exp(-z)$.

Here, however, we are interested in the general nature of the involved
functions, rather than particular physical model, so we wish to impose other
restrictions, namely the Liouvillian character of the solutions. For the
Whittaker equation the conditions for that are well known \cite{Morales} and
here they give
\begin{equation}
\begin{aligned}
    p &:= \mu+\kappa-\frac12 \in \mathbb{Z},\;\; \textrm{or}\\
    q &:= \mu-\kappa-\frac12 \in \mathbb{Z},
\end{aligned} 
\end{equation}
which, upon substitution, leads to
\begin{equation}
\begin{aligned}
    p+1 &= \frac{m}{\sqrt{m^2-E^2}}(\alpha+d)+\alpha \in \mathbb{Z},
    \;\; \textrm{or}\\
    q &= \frac{m}{\sqrt{m^2-E^2}}(\alpha+d)-\alpha \in \mathbb{Z}.
\end{aligned}
\end{equation}
In other words, when the potential is such that $\alpha=-d$ the solutions are
always Liouvillian, and otherwise the parameters must satisfy
\begin{equation}
    \frac{m}{\sqrt{m^2-E^2}} = \frac{l\pm\alpha}{d+\alpha},
    \;\; l\in \mathbb{Z},
\end{equation}
which in turn means that the integrable part of the spectrum is bounded by
\begin{equation}
    0 \leq \frac{E^2}{m^2} = 1 -\left(\frac{d+\alpha}{l\pm\alpha}\right)^2
    \leq 1.
\end{equation}
Note that when $\alpha$ is an integer itself, the case of $l=\pm\alpha$ is
excluded by $m\neq 0$.

For the particular case of $m^2=E^2$, Eq. \eqref{pre_whitt} is
effectively the Bessel equation
\begin{equation}
    w^2 \varphi''(w) + w \varphi'(w) + (w^2-(1-2\alpha)^2)\varphi(w)=0,
\end{equation}
when the variables are changed according to
\begin{equation}
    \psi(z) = \sqrt{z}\,\varphi(w),\;\;
    w^2 := -8m(\alpha+d)z,
\end{equation}
so, assuming $\alpha\neq d$, the solutions are given in term of the Bessel
functions
\begin{equation}
    \psi(z) = \sqrt{z} J_{1-2\alpha}(2\sqrt{-2m(\alpha+d)z}),
\end{equation}
and these are known to be Liouvillian only if their order, i.e. $1-2\alpha$,
is half of an odd integer.

Finally, for $\alpha = -d$ (and $m^2=E^2$), Eq. \eqref{pre_whitt} is solvable by
elementary functions of
\begin{equation}
    \psi(z) = c_1 z^{\alpha}+c_2 z^{1-\alpha}.
\end{equation}

\section{Summary}

To put the results together, we have shown that the one-dimensional Dirac
equation with a rational potential will lead to integrability conditions which
in general will be given by polynomials defined recursively. Also, that
rational potentials can be determined if some natural conditions are required
of the resulting equation itself. The direct and inverse results are as
follows.

The simplest form of $V = \alpha/z + m + \lambda z$ gives rise to the
biconfluent Heun equation (if none of the parameters is zero). Its solvability
is determined by the polynomial condition $p_{-1}(\alpha,m)=0$, obtained by one
of the relations \eqref{rec_rel} (together with the respective conditions on
$E$, $\lambda$ and $d$ given by \eqref{energ}) -- if any of the four can be
solved it gives a Liouvillian
solution. However, we are only interested in real values of the parameters and
the three theorems \ref{th1}, \ref{th2} and \ref{th3} characterize (to
some extent) the position and asymptotics of the real roots.

The special cases of the above potential, for vanishing parameters are
\begin{enumerate}
\item $\alpha=0$, the Dirac oscillator solvable by Hermite functions
\cite{dirac_poly}.
\item $m=0$, the Laguerre equation, solvable with
    $L_n^{\alpha-1/2}(\sqrt{-\lambda}z)$, with $4\lambda n=-E^2$.\\
    (These are Liouvillian functions when $n$ or $n+\alpha-1/2$ is an integer.)
\item $\lambda=0$, the Whittaker equation of section {\bf V}, summarized below.
\end{enumerate}

If we restrict the problem somewhat by requiring that $r(z)$ remains a
polynomial, although $V(z)$ might still be rational, we obtain a general
condition that the potential is of the form
\begin{equation}
    V(z) = \sum_{k=1}^{N} \frac{1}{z-c_k} + R_1(z) + R_2(z)\prod_{k=1}^N(z-c_k),
\end{equation}
with $R_1$ of degree $N-1$ such that
\begin{equation}
    R_1(c_k) + \sum_{\substack{l=1\\l\neq k}}^N \frac{1}{c_k-c_l} = 0,
\end{equation}
and $R_2$ and arbitrary polynomial. This is the most general rational potential
that leads to a polynomial coefficient $r(z)$. Due to $R_1$ being given only
implicitly (although uniquely) only the one-pole case in analyzed in section
{\bf IV}. The potential turns out to be
\begin{equation}
    V(z) = \frac{1}{z} + r_0 z + r_1 z^2,
\end{equation}
and the integrable values of the energy are $E^2 = 2r_0$.

Finally, choosing a predetermined $r(z)$ which is simpler than in the Heun
equation, but still has infinity as an irregular singular point and one other pole, 
leads to the Whittaker (confluent hypergeometric) equation. The
potential becomes
\begin{equation}
    V = m + \frac{\alpha}{z} + \frac{P'(z)}{P(z)},
    \label{potent}
\end{equation}
where $P$ is expressed by the associated Laguerre polynomial of degree $d$ as
\begin{equation}
    P(z) = L_d^{2\alpha-1}(-2m z),
\end{equation}
and $m$ is necessarily non-zero. The solutions are then given by the
Whittaker functions $W_{\kappa,\mu}(z)$,
and are additionally Liouvillian (can be expressed in closed form) when
\begin{enumerate}
    \item $\alpha = -d$ regardless of energy, or
    \item $2\alpha\in\frac12 +\mathbb{Z}$ for $E^2=m^2$,
    \item for particular levels satisfying 
    $\frac{m}{\sqrt{m^2-E^2}}(\alpha+d)\pm\alpha \in\mathbb{Z}$.
\end{enumerate}
The third possibility gives countably infinitely many energetic levels which
are confined to $E\in[-m,m]$. All the cases are put together
in table~\ref{tablica}.

\begin{table}[!h]
\begin{tabular}{|c|clclcl|}
\hline
\backslashbox{$\alpha$}{$E$} && $E\in\mathbb{C}$ &\phantom{dupa}&
$\frac{m}{\sqrt{m^2-E^2}}(\alpha+d)\pm\alpha\in\mathbb{Z}$&\phantom{dupa}&
$E^2=m^2$ \\ \hline
$\alpha\in\mathbb{C}$ &&
$W_{\kappa,\mu}(u),\,W_{-\kappa,\mu}(-u)$&& 
$W_{\kappa,\mu}(u),\,W_{-\kappa,\mu}(-u)\;\; [L]$&& 
$\sqrt{z} J_{1-2\alpha}(w),\, \sqrt{z} Y_{1-2\alpha}(w) $\\ 
$2\alpha-\frac12\in\mathbb{Z}$&&
$W_{\kappa,\mu}(u),\,W_{-\kappa,\mu}(-u)$&& 
$W_{\kappa,\mu}(u),\,W_{-\kappa,\mu}(-u)\;\; [L]$&& 
$\sqrt{z} J_{1-2\alpha}(w),\, \sqrt{z} Y_{1-2\alpha}(w) \;\; [L]$\\ 
$\alpha = -d$&&
$\sqrt{z} J_{\alpha-\frac12}(\frac{u}{2i}),\,
\sqrt{z} Y_{\alpha-\frac12}(\frac{u}{2i})\;\; [L]$&& 
$\sqrt{z} J_{\alpha-\frac12}(\frac{u}{2i}),\,
\sqrt{z} Y_{\alpha-\frac12}(\frac{u}{2i})\;\; [L]$&& 
$ z^{\alpha},\, z^{1-\alpha}\;\; [L]$\\ \hline
\end{tabular}
\caption{The list of bases of solutions depending on parameters. $W$ denotes
the Whittaker function, $J$ and $Y$ the Bessel functions. The auxiliary
variables are: $w=2\sqrt{-2m(\alpha+d)z}$, $u=2\sqrt{m^2-E^2}z$. $[L]$
signifies a Liouvillian pair of solutions.}
\label{tablica}
\end{table}

It should be clear from the general expression of the potential \eqref{potent}
that
it describes a very wide class, with an arbitrary number of singularities, thus
modeling drastically different physical situations. The mathematical analysis
carried out here gives the full information about the integrability and formal
solutions, but it is beyond its scope to select the appropriate eigenstates and
respective energies corresponding to a particular quantum system. However, just
as is the case with the Dirac oscillator, the integrability condition can be
thought of as some quantization rule. As shown in \cite{dirac_poly} dual rules
will hold for the vector coupling upon exchanging the roles of $E$ and $m$.
Either way, one can speak of some form of quasi-integrability here, since
Liouvillian solutions depend on values of (at least) two independent physical
parameters, so even before taking the physical restrictions into account, we
are left with at most countably many explicit solutions.

Concerning the property of shape invariance, the recovered potential contains
the known Coulomb case (albeit the spatial variable is then radial) listed in
\cite{Cooper}. This is the only case when the property holds, so that the whole
class is in fact new, and the method described can be used to uncover more
solvable cases by loosening some of the initial requirements. Future
generalizations might include target equations with more singular (regular or
not) points, or simply integrability conditions for given, more complex
potentials (trigonometric, meromorphic etc.).

Of course the Liouvillian solutions are not the only notion of integrability,
and most linear equations can be solved by means of series or numerically, but
the concept of an exact solution remains as the starting point in any such
approach. Any further study of the properties of the system in question,
especially qualitative, it greatly facilitated by working on explicit formulae,
which this work hopefully provides.

\acknowledgements
This research has been supported by grant No.
DEC-2011/02/A/ST1/00208 of National Science Centre of Poland.

\appendix
\section{}

When the degree $d=2n$ is even, $p_{-1}$ will be an
odd polynomial, so $m=0$ will make it trivially zero. Also, writing the
relation for even $l$ and taking the limit we get (in the first subcase)
\begin{equation}
    l(l+2\alpha-1)p_l = -2 (d-l+2)p_{l-2},
\end{equation}
so that only even indexed polynomials are generated from $p_d=1$, leading to
$p_0$ but not $p_{-1}$. It is thus necessary to consider polynomials
$r:=p_l/m$,  which changes the relation to
\begin{equation}
    l(l+2\alpha-1)r_l +2(l-1)p_{l-1} = -2(d-l+2)r_{l-2},
\end{equation}
Since $m$ does not enter the relation,
the limit of $m\rightarrow 0$ changes only the initial polynomial 
$r_{d-1} = -d$. Note that $r_l$ is defined only for odd $l$ and $p_{l-1}$
is the same as above, with its index even. The non-vanishing of energy forces
$d$ to be at least 2, by \eqref{energ}, and then $r_{-1} = 2\alpha/3$, so there is
one simple root (albeit excluded).

The $d\geq4$ case can be dealt with inductively. First the explicit formula for
the even indexed polynomials is
\begin{equation}
    p_l = \prod_{k=1+l/2}^{d/2} \frac{-2k(\alpha+k-1/2)}{d-2k+2}.
    \label{p_even1}
\end{equation}
Since $p_{d-2}=-d(2\alpha+d-1)/4$ has only one root, and as $l$ decreases,
there appear new simple roots at half (negative) integers, given by
\begin{equation}
    \alpha_k = \frac12-k,\;\;k=\frac{d}{2},\ldots,1+\frac{l}{2},
    \label{zero_seq}
\end{equation}
it follows that $p_l$ all share the first zero, and each has $(d-l)/2$ zeroes
altogether. The sign of the polynomial does not change past its final zero, and
can be determined from \eqref{p_even1}
\begin{equation}
    \mathrm{sgn}(p_l(\alpha_{l,M}+1/2)) = (-1)^{(d-l)/2},
    \label{sign_p}
\end{equation}
where the subscript in $\alpha_{l,M}$ indicates that the maximal
root depends on $l$, and the reason for taking the value specifically at
$\alpha_{l,M}+1/2$ will become clear in a moment.

The initial inductive step will be $l=d-1$ with the relation
\begin{equation}
    (d-1)(d+2\alpha-2)(-d)+2(d-2)p_{d-2} = -6 r_{d-3},
\end{equation}
where $r_{d-1}=-d$, and the other polynomials are linear in $\alpha$ but it
is more instructive not to substitute them explicitly. Now, at the first (and
only) root of $p_{d-2}$, $\alpha_{d/2}$, the above simply reduces to
\begin{equation}
    d(d-1) = - 6 r_{d-3}(\alpha_{d/2}) > 0,
\end{equation}
whereas at $\alpha_{d/2}+\frac12$ we have $(d+2\alpha_{d/2}-1)=0$ so
\begin{equation}
    2(d-2)p_{d-2}(\alpha_{d/2}+1/2) = -6 r_{d-2}(\alpha_{d/2}+1/2).
\end{equation}
By \eqref{sign_p}, the above translates to
\begin{equation}
    (-1)^{(d-d+2)/2} = -1 = - \mathrm{sgn}(r_{d-2}(\alpha_{d/2}+1/2)) = 
    -\mathrm{sgn}(r_{d-2}(\alpha_{d/2}+1)) < 0,
\end{equation}
where the last equality follows from the fact, that $r_{d-2}$ is linear
(by construction) and already changes sign once in
the interval $(\alpha_{d/2},\alpha_{d/2}+\frac12)$. Consequently there is a root, and
since $\alpha_{d/2}+1$ is the next zero in the sequence \eqref{zero_seq}, we know
the signs of $r_{d-3}$ at the first two $\alpha_k$.

The next step is to show that all following $r_l$ have the same signs at the
$(d-l+1)/2$ roots of $p_{l-1}$, in alternating sequence, and that there is an
additional change of sign at the $(d-l)/2$-th root of $p_{l-3}$. 
This is to extend the previous paragraph, where the signs agreed at
the first root, and there was a change at the next one. For $l<d-1$ the
relation at any zero $\alpha_k$ of $p_{l-1}$ is
\begin{equation}
    l(l-2k) r_l(\alpha_k) = -2(d-l+2)r_{l-2}(\alpha_k),
\end{equation}
so that the signs agree because $2+l<2k<d$. Thus as the signs alternate for
$r_l$ they also alternate for $r_{l-2}$ indicating $(d-l+1)/2$ simple zeroes.
Note that at the first zero $\alpha_{d/2}$, the signs of all $r_l$ are thus $-1$
and accordingly $\mathrm{sgn}(r_{l-2}(\alpha_{l-1,M})) = (-1)^{(d-l+1)/2}$.
As before we look what happens past the maximal zero at 
$\alpha_{l-1,M}+1/2=(1-l)/2$ so that the first term of the relation
is zero and 
\begin{equation}
    2(l-1)p_{l-1}(\alpha_{l-1,M}) = 
    -2(d-l+2)r_{l-2}(\alpha_{l-1,M}),
\end{equation}
or, in term of signs,
\begin{equation}
    (-1)^{(d-l+1)/2} = - \mathrm{sgn}(r_{l-2}(\alpha_{l-1,M}+1/2)).
\end{equation}
Because $r_{l-2}$ is a sum of polynomials of degree $(d-l+1)/2$, its degree
cannot be higher,
and the last change of sign means this is strictly the degree with all
roots real and simple. With no more sign changes it also follows that 
$r_{l-2}(\alpha_{l-1,M}+1) = (-1)^{(d-l+1)/2}$ and since that is the last root
of $p_{l-3}$ the last induction step is completed.

The last polynomial, $r_{-1}$, must thus have $d/2$ real simple zeroes, although
the last one is $\alpha=0$, for, as mentioned before, $p_{-1}$ is proportional to
$\alpha p_1$.

The third subcase is very similar to the first one, with positive $\lambda$
this time, and the even $p_l$ characterized by
\begin{gather}
    p_l = \prod_{k=1+l/2}^{d/2}\frac{k(2\alpha-2k-1)}{d-2k+2},\\
    \alpha_k = k+1/2,\\
    \mathrm{sgn}(p_l(\alpha_{l,m}-1/2)) = (-1)^{(d-l)/2},
    \label{p_even3}
\end{gather}
where $\alpha_{l,m}$ is the smallest root, whereas the greatest root is shared by all
$p_l$.  For odd $l$ the relation reads
\begin{equation}
    l(l+1-2\alpha) r_l -2lp_{l-1}=2(l-d-2)r_{l-2},
\end{equation}
and exactly as above the signs of $r_l$ and $r_{l-2}$ agree on zeroes of
$p_{l-1}$. The smallest root of $p_{l-1}$ is $1+l/2$ so the relation at
$\alpha_{l-1,m}-1/2$ gives
\begin{equation}
    l p_{l-1}(\alpha_{l-1,m}-1/2) = (d-l+2)r_{l-2}(\alpha_{l-1,m}-1/2).
\end{equation}
We can then proceed inductively, as before, by noting that $r_{d-1}=d$ so that
all $r_l$ are positive at the maximal root, and they alternate signs at
$\alpha_k$, with the additional change of sign resulting from the above
equation. The only trivial case to discard is $d=0$ for which $r_{-1}=1$, while
for $d\geq2$ it follows that there exist $d/2$ real simple roots.

Subcase 2 differs in two ways. Firstly, the sign of $\lambda$ is arbitrary, but
this only means that for $\lambda<0$ the signs of subsequent $r_l$ will be
opposite at zeroes
of $p_l$ instead of the same. That will not change the fact that the signs
alternate and account for all but one zero. Here is where the second difference
appears, for we have
\begin{gather}
    p_l = \prod_{k=1+l/2}^{d/2}\frac{k(2\alpha+2k-1)}{2k-d-2},\\
    \alpha_k = 1/2 - k,
    \label{p_even2}
\end{gather}
and as the relation is now
\begin{equation}
    l(l+2\alpha-1)r_l-2(l+2\alpha-1)p_{l-1}=2\lambda(l-d-2)r_{l-2},
\end{equation}
it means that at $\alpha_{l-1,M}+1/2$ we get
\begin{equation}
    0 = 2\lambda(l-d-2)r_{l-2}((1-l)/2).
\end{equation}
So that the last zero of $r_l$ is known explicitly and it ensures the desired
changes of signs at roots of the appropriate $p_l$. Since $r_{d-1}= 2\alpha+d$
and $p_{d-2}=-d(2\alpha+d-1)/2$ have different roots (and hence the sign
changes accordingly), induction on $l$ yields the existence of $1+d/2$ roots of
$r_{-1}$ with the explicit root being $\alpha=0$. Note that this also holds for
$d=0$ where $r_{-1}=2\alpha$.

Finally, in subcase 4, there is a complication due to $r_{d-1}=-2\alpha+d+1$, 
$p_{d-2}= d(-2\alpha+d+1)/4$, so that all polynomials share a root. As
mentioned before, it corresponds to the degenerate $E=0$ case. This is the
greatest of the roots of $p_l$ which are
\begin{gather}
    p_l = \prod_{k=1+l/2}^{d/2}\frac{k(-2\alpha+2k+1)}{\lambda(d-2k+2)},\\
    \alpha_k = k+1/2.
    \label{p_even4}
\end{gather}
Let us consider $\lambda>0$ for clarity (the negative case is completely
analogous). Note that $p_l$ will be positive for $\alpha$ less than the
smallest zero because the leading coefficient will be $(-\alpha)^{(d-l)/2}$
and since the relation is
\begin{equation}
    l(l+1-2\alpha)r_l +2(l-2\alpha)p_{l-1}=2(d-l+2)r_{l-2},
    \label{rel4}
\end{equation}
the same will be true for $r_l$ and its smallest zero. We will see that apart
from $\alpha=(d+1)/2$ all the zeroes of $r_{l-2}$ lie between the zeroes of
$p_{l-3}$. Direct computation shows that the smaller zero of $r_{d-3}$ is
$3d(d-1)/(6d-4)$ which lies between $(d-1)/2$ and $(d+1)/2$. In particular
$r_{d-3}$ is positive at $(d-1)/2 = \alpha_{d-4,m}$. Assume that this is true
for $r_l$ and the relation for zeroes of $p_{l-1}$ other than $(d+1)/2$ gives
\begin{equation}
    l(l+1-2\alpha_k)r_l(\alpha_k) = 2(d-l+2)r_{l-2}(\alpha_k),
\end{equation}
with the coefficient of $r_l$ being negative due to \eqref{p_even4}. At
$\alpha_{l-1,m}-1/2=(l+1)/2$ the relation reads
\begin{equation}
    -2p_{l-1}(\alpha_{l-1,m}-1/2) = 2(d-l+2)r_{l-2}(a_{l-1,m}-1/2) < 0,
\end{equation}
but at $\alpha_{l-3,m}=l/2$ we get
\begin{equation}
    l r_l(\alpha_{l-1,m}-1) = 2(d-l+2)r_l(\alpha_{l-1,m}-1) > 0,
\end{equation}
because $r_l$ is positive to the left of its smallest zero, which is greater
than $\alpha_{l-1,m}$. In other words, $r_{l-2}$ is positive at
$\alpha_{l-3,m}$, and negative at $\alpha_{l-1,m}$ and then changes the sign 
between the  subsequent zeros of $p_{l-1}$ because $r_l$ does so. At the last
but one zero (of $p_{l-1}$) it thus has the sign of $(-1)^{(d-l-1)/2}$ (number
of zeroes of $p_{l-1}$ except the last), and close to the left of the last
zero $p_{l-1}$ and $r_l$ both have the sign $(-1)^{(d-l-1)/2}$ because their
leading terms are $(-\alpha)^{(d-l+1)/2}$. However, they enter the relation
\eqref{rel4} with negative coefficients ($l+1-2\alpha<0$ when $\alpha>d/2$)
which means that $r_{l-2}$ changes sign additionally before the last root of
$(d+1)/2$. Last thing to check is the $d=0$ case, for which 
$r_{-1}= 1-2\alpha$. This finishes the proof that $r_{-1}$ has $1+d/2$ real
simple roots.



\begin{thebibliography}{9}


\bibitem{Martin} A.~Bermudez, M.~A.~Martin-Delgado, E.~Solano,
Phys. Rev. A {\bf76}, 041801 (R) (2007).

\bibitem{Primitivo} Acosta-Hum\'anez~P.~B., PhD Thesis, Universitat
Polit\`ecnica de Catalunya (2009). arXiv:0906.3532

\bibitem{Cooper} Cooper~F., Khare~A. and Sukhatme~U.,
Phys. Repts. {\bf 251} 267--385 (1995).

\bibitem{Prim2} Acosta-Hum\'anez~P.~B. and Bl\'azquez-Sanz~D.,
Discrete Contin. Dynam. Syst. Series B {\bf 10}, 2\&3, 265--293 (2008).

\bibitem{Prim3} Acosta-Hum\'anez~P.~B., Morales-Ruiz~J. and Weil~J.~A.,
    Rep. Math. Phys. {\bf 67}(3), 305--374, (2011).

\bibitem{TS} Stachowiak~T., Szyd\lpb owski~M. and Maciejewski~A.~J.,
J. Math. Phys. {\bf 47} 032502 (2006).

\bibitem{dirac_poly} Stachowiak~T.,
J. Math. Phys. {\bf 52}, 012301 (2011).

\bibitem{Kaplansky}
Kaplansky~I., ``An Introduction to differential algebra'', Hermann, Paris
(1957).

\bibitem{Kovacic}
Kovacic~J., J.~Symbolic~Comput. {\bf 2}(1): 3--34 (1986).

\bibitem{Duval}
Duval~A. and Loday-Richaud~M., Appl. Algebra Engrg. Comm. Comput. 
{\bf 3}(3), 211--246, (1992).

\bibitem{Morales}
Juan J.~Morales-Ruiz,
``Differential Galois theory and non-integrability of Hamiltonian
systems'', (Birkh\"auser Verlag, Basel, 1999).

\end{thebibliography}
\end{document}